\documentclass[journal]{IEEEtran}
\newcommand{\papertitle}{Noise Analysis of Open-Loop and Closed-Loop SAW Magnetic Field Sensor Systems}

\usepackage[normalem]{ulem}
\usepackage{cancel}

\usepackage{cite}

\ifCLASSINFOpdf
  \usepackage[pdftex]{graphicx}
\else
\fi
\usepackage{amsmath}
\usepackage{amssymb}
\usepackage{sistyle}

\usepackage{hyperref}
\hypersetup{
    unicode=false,
    pdftoolbar=true,
    pdfmenubar=true,
    pdffitwindow=false,
    pdfstartview={FitH},
    pdftitle={\papertitle},
    pdfauthor={Durdaut et al.},
    pdfsubject={\papertitle},
    pdfcreator={Durdaut et al.},
    pdfproducer={Kiel University},
    pdfnewwindow=true,
    colorlinks=true,
    linkcolor=black,
    citecolor=black,
    filecolor=black,
    urlcolor=black
}

\usepackage{graphicx}
\usepackage{caption}
\usepackage{subcaption}

\usepackage{epstopdf}

\usepackage{upgreek}

\usepackage{cancel}

\usepackage{booktabs}

\usepackage{multirow}

\usepackage{mathtools}

\usepackage{cuted}
\usepackage{mathrsfs}

\usepackage{array}
\begin{document}
\bstctlcite{IEEEexample:BSTcontrol}
%
\title{\papertitle}
%
%
%

\author{Phillip~Durdaut, \textit{Student Member, IEEE},
        Anne~Kittmann,
				Enrico~Rubiola, \textit{Member, IEEE},
				Jean-Michel~Friedt,
				Eckhard~Quandt,
        Reinhard~Kn\"ochel, \textit{Life Fellow, IEEE},
        and Michael~H\"oft, \textit{Senior Member, IEEE}\\
\thanks{P.~Durdaut, R.~Kn\"ochel and M.~H\"oft are with the Chair of Microwave Engineering, Institute of Electrical Engineering and Information Technology, Faculty of Engineering, Kiel University, Kaiserstr. 2, 24143 Kiel, Germany.}
\thanks{A.~Kittmann and E.~Quandt are with the Chair of Inorganic Functional Materials, Institute for Materials Science, Faculty of Engineering, Kiel University, Kaiserstr. 2, 24143 Kiel, Germany}
\thanks{E.~Rubiola and J.-M.~Friedt are with the FEMTO-ST Institute, Department of Time and Frequency, Universit\'{e} de Bourgogne Franche-Comt\'{e} (UBFC), and CNRS, ENSMM, 26 Rue de l'\'{E}pitaphe, 25000 Besan\c{c}on, France. E.~Rubiola is also with the Physics Metrology Division, Istituto Nazionale di Ricerca Metrologica (INRiM), Strada Delle Cacce 91, 10135 Torino, Italy.}}

%
%

\markboth{Durdaut \MakeLowercase{\textit{et al.}}: \papertitle}
{}
%



\maketitle

\begin{abstract}
Transmission surface acoustic wave (SAW) sensors are widely used in various fields of application. In order to maximize the limit of detection (LOD) of such sensor systems, it is of high importance to understand and to be able to quantify the relevant noise sources. In this paper, low noise readout systems for the application with a SAW delay line magnetic field sensor in an open-loop and closed-loop configuration are presented and analyzed with regard to their phase noise contribution. By applying oscillator phase noise theory to closed-loop sensor systems, it is shown that the phase noise of the SAW delay line oscillator can be predicted accurately. This allows the derivation of expressions for the limits of detection for both readout structures. Based on these equations, the equivalence between the LOD of open-loop and closed-loop SAW delay line readout can be shown analytically, assuming that the sensor contributes the dominant phase noise. This equality is verified by measurements. These results are applicable to all kinds of phase sensitive delay line sensors.
\end{abstract}


%
\IEEEpeerreviewmaketitle

%
%
%
%


\section{Introduction}
\label{sec:introduction}
Surface acoustic wave (SAW) devices started to become attractive with the invention of the interdigital transducer (IDT) in 1965 \cite{White.1965} which allows to excite SAWs on piezoelectric substrates in an efficient way. Advantageous properties like small size, low cost, and high reproducibility \cite{Ruppel.1993} make SAW technology very attractive for sensor applications \cite{White.1985,Liu.2016}. Among many others, SAW sensors for measuring temperature \cite{Neumeister.1989,Hauden.1981}, pressure \cite{Scherr.1996,Jungwirth.2002}, magnetic fields \cite{Webb.1979,Smole.2003,Kadota.2011,Li.2012,Bocchiaro.2013,Elhosni.2014a,Wang.2015,Elhosni.2015,Wang.2016,Mishra.2017,Tong.2017,Polewczyk.2017,Kittmann.2018,Liu.2018,Wang.2018a}, humidity \cite{Caliendo.1993}, and vibration \cite{Filipiak.2011} or for the detection of gases \cite{Devkota.2016} and biorelevant molecules \cite{Laenge.2008,Hirst.2008}, respectively, have been reported.

A SAW is excited by applying an electrical field on an IDT that is patterned on a piezoelectric material. The resulting mechanical wave propagates perpendicular to the direction of the IDT in both directions on the surface of the piezoelectric substrate with typical wave velocities between \SI{3000}{m/s} and \SI{5000}{m/s} \cite{Fischerauer.2008,Liu.2016}. For sensing applications the substrate's surface is frequently coated with an additional layer which reacts to changes of the physical quantity to be measured, and in turn alters the SAW in its amplitude and in its velocity.

Throughout this paper, we solely focus on a two-port delay line sensor that comprises two IDT electrodes placed in some distance to each other. A sensing layer between the IDTs attenuates and affects the velocity of the propagating wave in dependence of the physical quantity to be measured. Because a surface acoustic wave is traveling about five orders of magnitude slower compared to the electromagnetic propagation the delay line results in typical group delays $\tau_{\mathrm{g}}$ from several hundred nanoseconds to several microseconds, depending on the SAW device's size.

There are two different schemes for the readout of SAW delay line sensors. A straightforward approach is to compare the phase of the sensor's output signal with a reference phase \cite{Fischerauer.2008} in an open-loop system, thus creating a delay line phase discriminator \cite{Sullivan.1990,Schiek.2006}. Alternatively, readout of such sensors can be achieved with a closed-loop approach by including the sensor into the feedback loop of an oscillator, thus changing oscillation frequency with the sensing function \cite{Lewis.1974,Parker.1982,Parker.1988}. Both approaches might suffer from noise which is introduced into the sensor system by the required electrical components. However, even if all the electronic components are designed very carefully such that their noise contributions can be neglected, one crucial question remains: Is one of these approaches superior to the other regarding the overall achievable limit of detection (LOD)? Recently, we addressed this question in a theoretic study \cite{Durdaut.2019}. In this paper, a SAW magnetic field sensor is operated in open-loop configuration and in a closed-loop system. Based on phase noise measurements of the individual electronic components of both systems, the earlier presented expressions for the phase noise and the LOD are verified.

This paper is organized as follows: Sec.~\ref{sec:saw_sensor} briefly introduces the SAW magnetic field sensor utilized in this investigation. Based on the sensor-specific requirements open-loop and closed-loop readout systems are presented in Sec.~\ref{sec:readout_systems}. A comparison regarding the readout system's overall phase noise is presented in Sec.~\ref{sec:noise_analysis} based on the individual phase noise contributions of both the electrical components and the sensor-intrinsic noise. This article ends with a summary in Sec.~\ref{sec:conclusion}.

\section{SAW Sensor}
\label{sec:saw_sensor}
The utilized SAW delay line sensor has been produced in the Kiel Nanolaboratory and is based on a \SI{500}{\mu m} thick ST-cut quartz substrate. A delay line is formed using two split-finger IDT electrodes with 25 finger pairs, a periodicity of \SI{28}{\mu m} and a finger width of \SI{3.5}{\mu m} with an IDT center-to-center length of ${L = \SI{4.64}{mm}}$. A SiO$_2$ layer with a thickness of \SI{4.5}{\mu m} deposited on top of the IDTs and the delay line acts as a guiding layer for the surface acoustic Love wave. A magnetostrictive material (Fe$_{90}$Co$_{10}$)$_{78}$Si$_{12}$B$_{10}$ with a thickness of \SI{200}{nm} and a length of \SI{3.8}{mm} is deposited on top of the guiding layer and between the IDTs. Further details about the fabrication can be found in \cite{Kittmann.2018}.

The sensor principle \cite{Kittmann.2018} is based on the magnetoelastic effect which changes the shear modulus of the magnetostrictive material as a function of the material's magnetization, i.e. by a surrounding magnetic flux density $B$. Due to the relation between the elastic properties and the wave's propagation velocity \cite{Smole.2003}, the phase angle $\varphi$ of the sensor signal changes with the magnetic flux density $B$.

For this type of sensor, not only the expected dependence of the magnetic sensitivity $\mathcal{S}_{\mathrm{mag}}$ on a DC magnetic bias flux density $B_{\mathrm{bias}}$ exists. Also strong interdependencies between the magnetic insertion loss $\mathrm{IL}_{\mathrm{mag}}$, the sensor's intrinsic phase noise and both the SAW input power $P_{\mathrm{SAW}}$ and $B_{\mathrm{bias}}$ are observed. Further details about the magnetic properties as well as the noise behavior of this kind of sensor will be published in the near future. Generally, an additional DC bias flux density $B_{\mathrm{bias}}$ can be applied to optimize the phase noise behavior and thus, the limit of detection (LOD). However, this requires the sensor to be surrounded by an additional coil fed by a current from an ultra-low noise current source which is also capable of providing relatively high currents in order to perform a certain magnetic saturation prior to the measurements. Yet, the sensor used in this investigation is quite suited to operate without any DC bias flux density. For simplification, and because the focus in this article is on the readout systems, the magnetic operating point is set to $B_{\mathrm{bias}} = 0$. Tab. \ref{table:properties_saw_sensor} summarizes the most important electrical and magnetic properties of the SAW magnetic field sensor in its operating point. 

\begin{table}[!t]
\renewcommand{\arraystretch}{1.3}
\caption{Summary of the electrical and magnetic properties of the utilized magnetic field SAW sensor in its operating point.}
\label{table:properties_saw_sensor}
\centering
\begin{tabular}{|c|c|}
\hline
Length of the delay line & $L = \SI{4.64}{mm}$\\
Acoustic wavelength & $\lambda = \SI{28}{\mu m}$\\
\hline
Center frequency & $f_0 = \SI{144.8}{MHz}$\\
\SI{-3}{dB} bandwidth & $\Delta f = \SI{4.6}{MHz}$\\
Return loss & $\mathrm{RL} > \SI{20}{dB}$\\
Insertion loss of the delay line & $\mathrm{IL}_{\mathrm{elec}} = \SI{20}{dB}$\\
Magnetic loss & $\mathrm{IL}_{\mathrm{mag}} = \SI{8}{dB}$\\
Overall insertion loss & $\mathrm{IL} = \mathrm{IL}_{\mathrm{elec}}+\mathrm{IL}_{\mathrm{mag}} = \SI{28}{dB}$\\
\hline
Phase velocity & $v_{\varphi} = f_0 \lambda = \SI{4054}{m/s}$\\
Phase delay & $\tau_{\varphi} = L / v_{\mathrm{\varphi}} = \SI{1145}{ns}$\\
Phase slope of the delay line & $\partial \varphi / \partial f = \SI{-8.73}{rad/MHz}$\\
Group delay & $\tau_{\mathrm{g}} = -\partial \varphi / ( \partial f ~ 2 \pi) = \SI{1389}{ns}$\\
Group velocity & $v_{\mathrm{g}} = L / \tau_{\mathrm{g}} = \SI{3341}{m/s}$\\
\hline
Electrical sensitivity & $\mathcal{S}_{\mathrm{elec}} = 2 \pi \tau_{\mathrm{g}} = \SI{8.73}{rad/MHz}$\\
Magnetic sensitivity & $\mathcal{S}_{\mathrm{mag}} = \partial \varphi / \partial B = \SI{16.5}{rad/mT}$\\
\hline
SAW input power & $P_{\mathrm{SAW}} = \SI{0}{dBm}$\\
Magnetic operating point & $B_{\mathrm{bias}} = 0$\\
\hline
Sensor's phase noise $S_{\varphi}(f)$ & See Fig.~\ref{fig:saw_phasenoise}\\
\hline
\end{tabular}
\end{table}

\section{Readout Systems}
\label{sec:readout_systems}
\begin{figure*}[ht!]
	~
	\begin{subfigure}[t]{1\textwidth}
		\centering
		\includegraphics[width=0.75\linewidth]{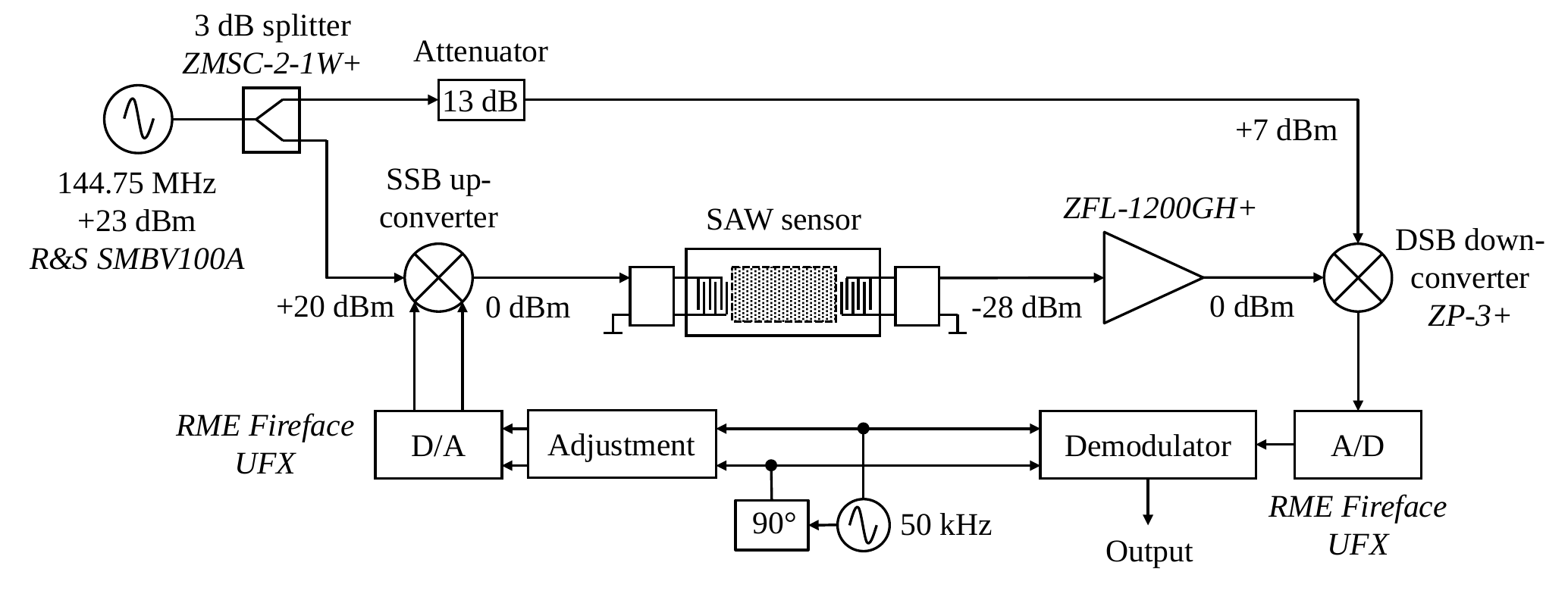}
		\caption{Open-loop readout system with inherent phase noise suppression of the high-frequency local oscillator. A low phase noise signal of a numerically controlled oscillator with a frequency of \SI{50}{kHz} is transposed to the SAW sensor's operating frequency $f_0 = \SI{144.8}{MHz}$ and received using a floating local oscillator, which largely suppresses its phase noise. Phase detection is performed in the digital domain after analog-to-digital conversion.}
		\label{fig:open_loop_system}
	\end{subfigure}
	~
	\begin{subfigure}[t]{1\textwidth}
		\centering
		\includegraphics[width=0.75\linewidth]{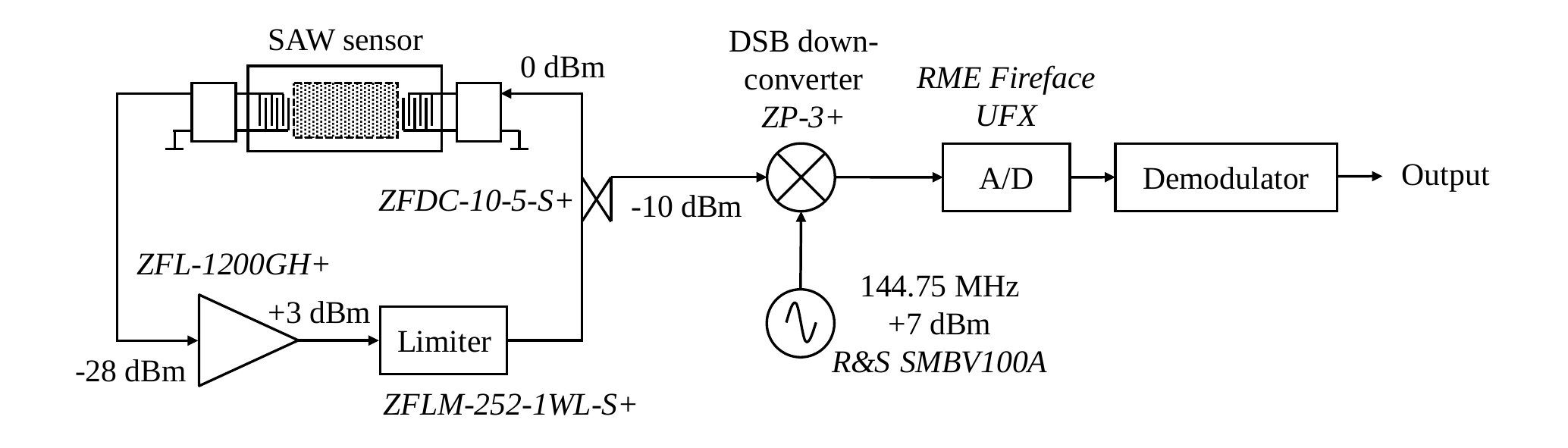}
		\caption{Closed-loop readout system with the SAW sensor included in the feedback loop of an amplifier leading to an oscillation with a frequency of approx. $f_0 = \SI{144.8}{MHz}$. A limiter is used to stabilize the SAW sensor's input power to \SI{0}{dBm}. Frequency detection is performed in the digital domain after downconverting the oscillation signal to an intermediate frequency of \SI{50}{kHz} and subsequent analog-to-digital conversion.}
		\label{fig:closed_loop_system}
	\end{subfigure}
	~
	\caption{Developed open- and closed-loop readout systems for the determination of the applied magnetic field. In the open-loop systems a magnetic flux density to be measured leads to a phase modulation of the SAW sensor's output signal whereas the same leads to a frequency modulation in the closed-loop oscillator.}
	\label{fig:readout_systems}
\end{figure*}

The developed open- and closed-loop readout systems are depicted in Fig.~\ref{fig:readout_systems} together with the according power levels. Both systems are mostly based on RF components by \textit{Mini-Circuits} with all their identifiers given in the system diagrams. In order to allow a comparison between both methods care has been taken to feed the SAW sensor in each case with the same input power of ${P_{\mathrm{SAW}} = \SI{0}{dBm}}$ because the SAW sensor's intrinsic phase noise is a function of $P_{\mathrm{SAW}}$.

\subsection{Open-Loop System}
\label{subsec:readout_systems_open_loop_system}

\begin{figure}[t]
	\centering	
	\begin{subfigure}[t]{0.5\textwidth}
		\centering
		\includegraphics[width=1.0\linewidth]{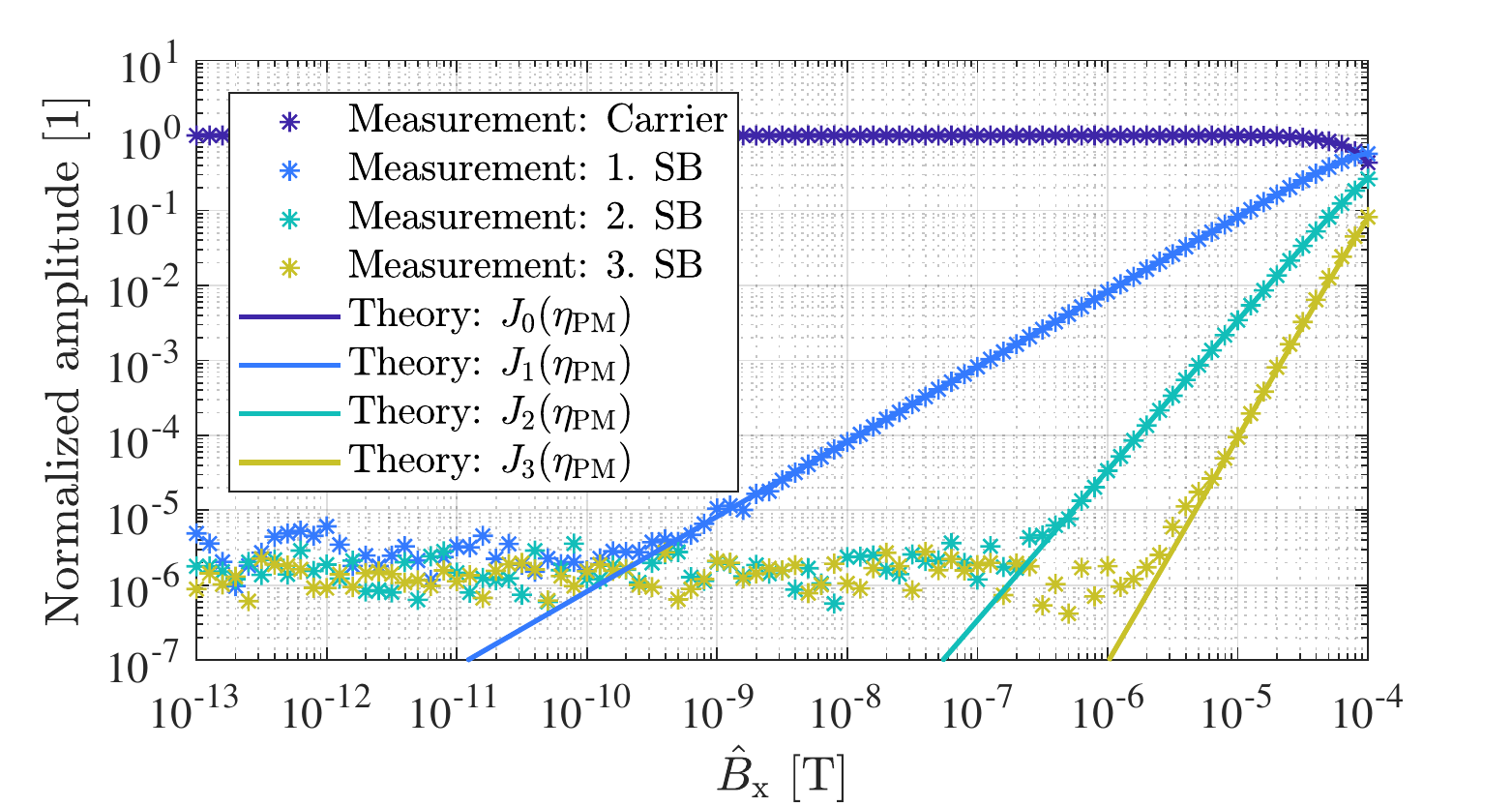}
		\caption{Phase modulated signal in the open-loop system}		
		\label{fig:pm_signal}
	\end{subfigure}
	~
	\begin{subfigure}[t]{0.5\textwidth}
		\centering
		\includegraphics[width=1.0\linewidth]{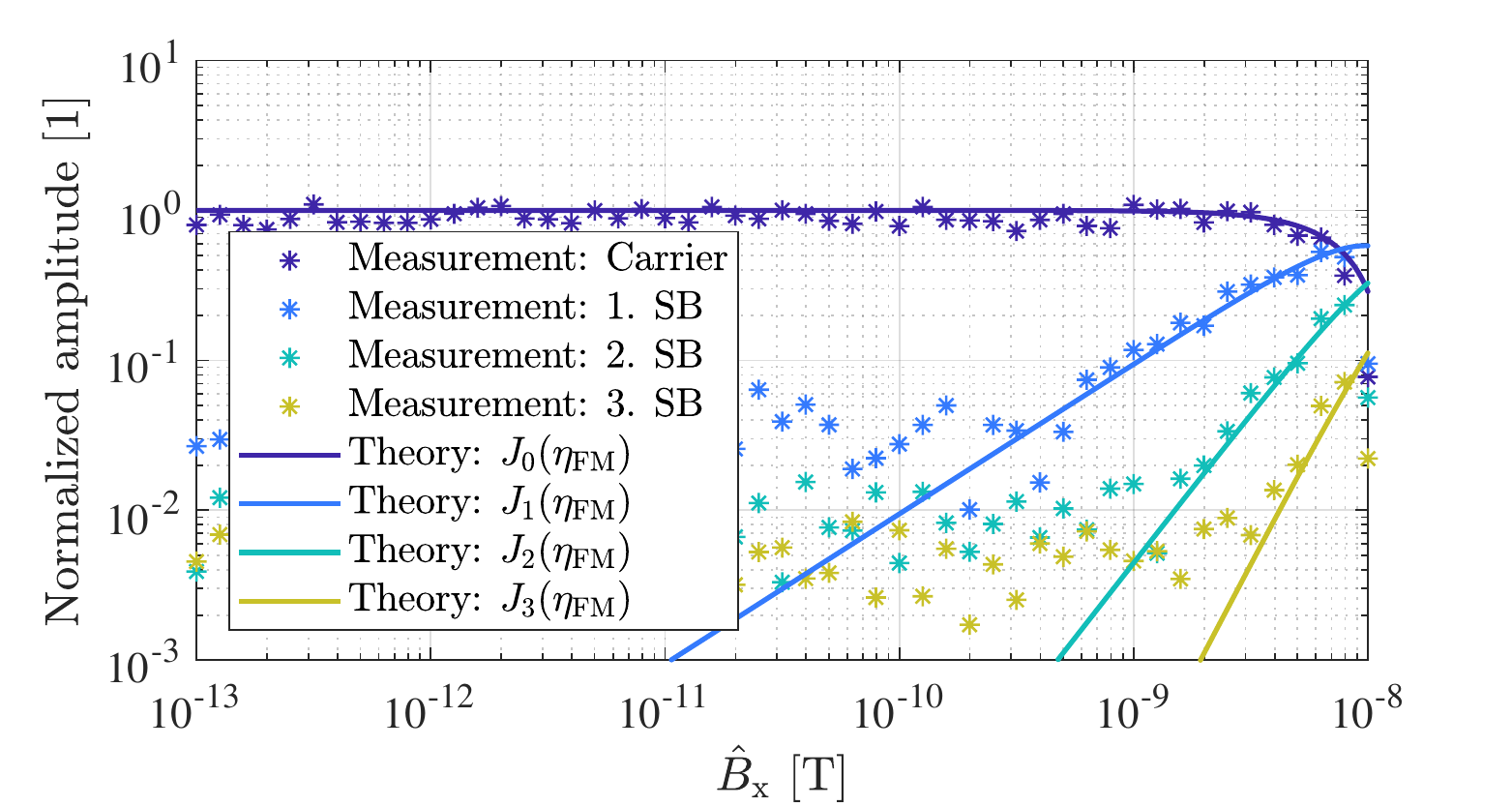}
		\caption{Frequency modulated signal in the closed-loop system}
		\label{fig:fm_signal}
	\end{subfigure}
	~
	\caption{Verification of the modulation purities for both sensor systems (Fig.~\ref{fig:readout_systems}) by measured carrier and sideband (SB) amplitudes and comparison with the theoretical expectations. The measurements were conducted for a frequency ${f_{\mathrm{x}} = \SI{10}{Hz}}$ with the sensor in its operating point according to Tab.~\ref{table:properties_saw_sensor}.}
	\label{fig:pm_fm_signal}
\end{figure}

The open-loop readout system (Fig.~\ref{fig:open_loop_system}) is based on a heterodyne structure in which the output signal of a phase stable numerically controlled oscillator (NCO) with a frequency of \SI{50}{kHz} is upconverted to the SAW device's passband at ${f_0 = \SI{144.8}{MHz}}$ by means of a single sideband (SSB) upconverter and a local oscillator (LO). The SSB upconverter suppresses the undesired lower sideband which would also fall into the passband of the SAW device and which represents the image frequency with regard to the subsequent downconverter. Sufficient sideband suppression (typically about \SI{-60}{dB}) is achieved by adjusting both amplitude and phase of the SSB converter drive signals numerically. The SSB upconverter consists of a 2-way-90$^\circ$ power splitter \textit{ZMSCQ-2-180+} at the input and a 2-way-0$^\circ$ combiner \textit{ZMSC-2-1W+} at the output. In-between, two level-17 mixers \textit{ZAD-1H+} perform the frequency conversion. The sensor's output signal is amplified and downconverted to the original frequency of \SI{50}{kHz} utilizing a double sideband (DSB) mixer using the same LO. Thus, the phase noise of the high-frequency LO is largely suppressed. In fact, the degree of phase noise suppression depends on the group delay $\tau_{\mathrm{g}}$ of the SAW sensor and on the offset frequency \cite{Durdaut.2018}. Assuming a sinusoidal magnetic flux density to be measured ${B_{\mathrm{x}}(t) = \hat{B}_{\mathrm{x}} \cos\left( 2 \pi f_{\mathrm{x}} t \right)}$ the sensor's output signal 
\begin{align}
		s_{\mathrm{PM}}(t) \propto \cos\big( 2 \pi f_0 t + \mathcal{S}_{\mathrm{PM}} \hat{B}_{\mathrm{x}} \cos\left( 2 \pi f_{\mathrm{x}} t \right) + \psi_{\mathrm{OL}}(t)\big)
		\label{eqn:sPMt}
\end{align}
is phase modulated (PM) by $B_{\mathrm{x}}(t)$ with a sensitivity of ${\mathcal{S}_{\mathrm{PM}} = \mathcal{S}_{\mathrm{mag}} = \SI{16.5}{rad/mT}~(\approx~\SI{945}{^\circ/mT})}$. From basic modulation theory, it is well-known that the modulation index ${\eta_{\mathrm{PM}} = \mathcal{S}_{\mathrm{PM}} \hat{B}_{\mathrm{x}}}$ is linked to the carrier-to-sideband ratios of phase-modulated signals by \textit{Bessel functions} of the first kind ${J_{\nu}(\eta_{\mathrm{PM}})}$, where $\nu$ is the number of the sideband (SB) \cite[p. 141 ff.]{Ziemer.2009}. Based on this relationship, prior to the noise measurements, the modulation purity and thus, the sensitivity $\mathcal{S}_{\mathrm{PM}}$ is verified by a series of measurements for various amplitudes $\hat{B}_{\mathrm{x}}$ of the signal to be measured with the results shown in Fig.~\ref{fig:pm_signal}.
In Eq.~\eqref{eqn:sPMt} $\psi_{\mathrm{OL}}(t)$ describes the overall phase noise of the open-loop system, i.e. the phase noise of the SAW sensor and the phase noise of the electronic readout system. Both, the contribution of the sensor as well as the individual contributions of the various electrical components will be analyzed in Sec.~\ref{sec:noise_analysis}.

\subsection{Closed-Loop System}
\label{subsec:readout_systems_closed_loop_system}

The structure of the closed-loop readout system is shown in Fig.~\ref{fig:closed_loop_system}. The SAW sensor is included in the feedback branch of an amplifier which compensates for the sensor's insertion loss, and thus leads to an oscillation at approx. ${f_0 = \SI{144.8}{MHz}}$. An additional limiter is used to stabilize the SAW sensor's input power to \SI{0}{dBm}. Because of the sensor's linear phase response over a wide frequency range with a slope of ${\partial \varphi / \partial f = \SI{-8.73}{rad/MHz}}$, oscillations with a frequency spacing of ${\delta f_0 = 1/\tau_{\mathrm{g}} \approx \SI{720}{kHz} < \Delta f}$ are generally possible. Therefore, an additional phase shifter is not needed. Utilizing a \SI{-10}{dB} directional coupler the oscillator signal is coupled into a mixer for downconversion to an intermediate frequency of \SI{50}{kHz} in order to perform the subsequent analog-to-digital conversion and the demodulation. Assuming again a sinusoidal magnetic flux density to be measured ${B_{\mathrm{x}}(t) = \hat{B}_{\mathrm{x}} \cos\left( 2 \pi f_{\mathrm{x}} t \right)}$ the oscillator signal 
\begin{align}
		s_{\mathrm{FM}}(t) \propto \cos\Big( 2 \pi f_0 t + \frac{\mathcal{S}_{\mathrm{FM}} \hat{B}_{\mathrm{x}}}{f_{\mathrm{x}}} \sin\left( 2 \pi f_{\mathrm{x}} t \right) + \psi_{\mathrm{CL}}(t)\Big)
		\label{eqn:sFMt}
\end{align}
is frequency modulated (FM) by $B_{\mathrm{x}}(t)$ with a sensitivity of ${\mathcal{S}_{\mathrm{FM}} = \mathcal{S}_{\mathrm{mag}}/\mathcal{S}_{\mathrm{elec}} = \SI{1.9}{MHz/mT}}$ or with a modulation index of ${\eta_{\mathrm{FM}} = \mathcal{S}_{\mathrm{FM}} \hat{B}_{\mathrm{x}} / f_{\mathrm{x}}}$, respectively. As already shown for the open-loop system, the FM purity and the sensitivity $\mathcal{S}_{\mathrm{FM}}$ is verified by measurement of the carrier-to-sideband ratios and subsequent comparison with the theoretical expectations according to the \textit{Bessel functions} of the first kind ${J_{\nu}(\eta_{\mathrm{FM}})}$. The results are shown in Fig.~\ref{fig:fm_signal} in which the variance of the measured amplitudes is distinctly higher as for the PM case because of the high phase noise in the closed-loop-system.
In Eq.~\eqref{eqn:sFMt} $\psi_{\mathrm{CL}}(t)$ describes the overall oscillator phase noise of the closed-loop system, i.e. the phase noise due to the SAW sensor and due to the electronic readout system, both further discussed in Sec.~\ref{sec:noise_analysis}.

\section{Noise Analysis}
\label{sec:noise_analysis}
\begin{figure}[t]
	\centering
	\includegraphics[width=0.5\textwidth]{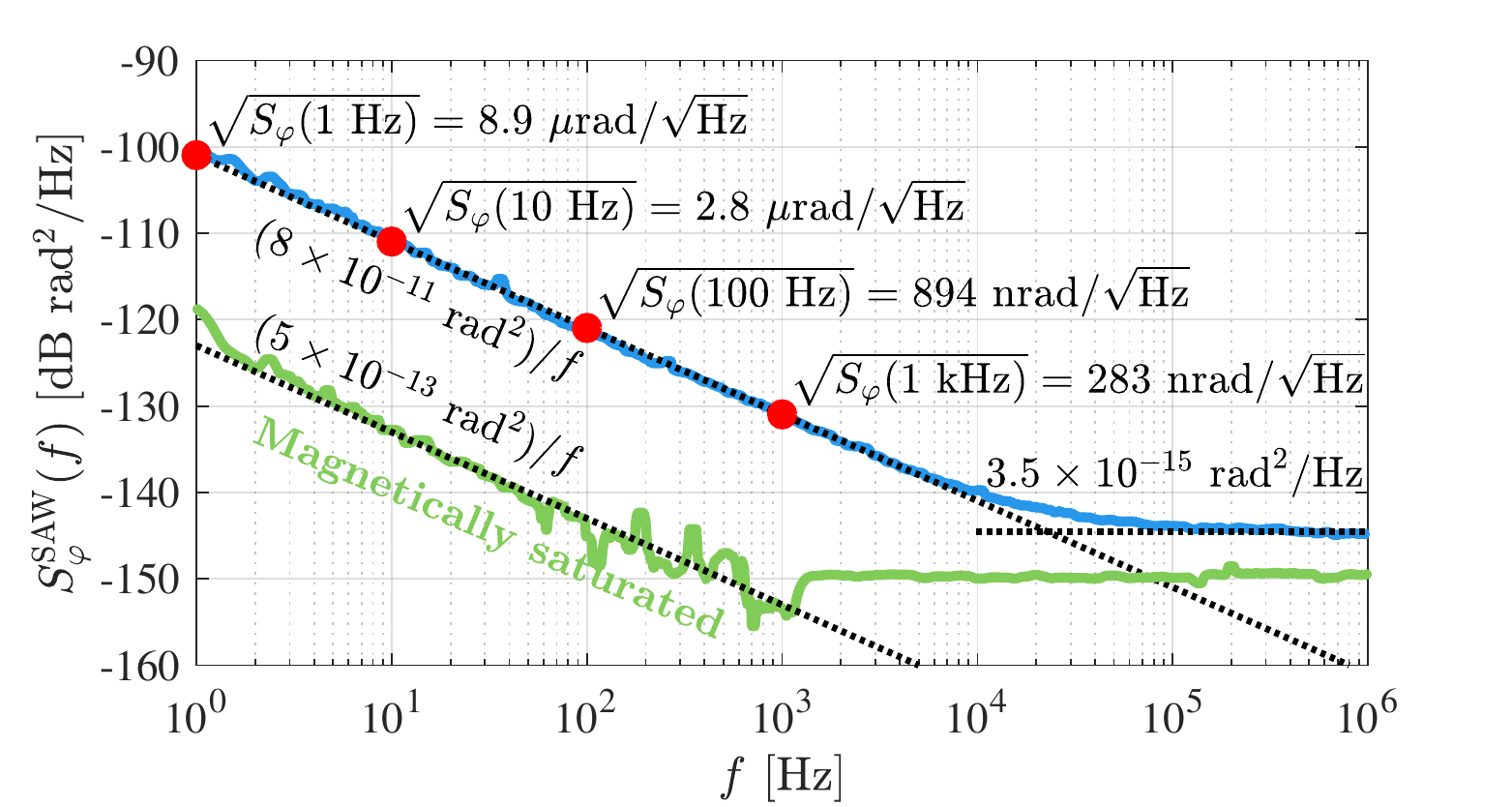}
	\caption{Measured power spectral density of the random phase fluctuations of the SAW sensor in its operating point according to Tab.~\ref{table:properties_saw_sensor} (blue) and under magnetic saturation (green).}
	\label{fig:saw_phasenoise}
\end{figure}

Random fluctuations of an arbitrary phase $\varphi(t)$ are best described by the one-sided power spectral density of the random phase fluctuations $S_{\varphi}(f)$ in units of $\mathrm{rad}^2/\mathrm{Hz}$. For historical reasons, phase noise analyzers give the results in terms of a phase noise density spectrum $\mathscr{L}(f)$ which is defined as ${\mathscr{L}(f) = 1/2~S_{\varphi}(f)}$ and usually given in units of $\mathrm{dBc}/\mathrm{Hz}$ \cite{Rubiola.2000,IEEE.2009}. In order to simplify mathematical expressions and to stay with SI units, $S_{\varphi}(f)$ is used throughout this article. The logarithmic representation $10 \log_{10}(S_{\varphi}(f))$ is then given in units of $\mathrm{dB}~\mathrm{rad}^2/\mathrm{Hz}$. A useful model for describing the frequency dependence of a power spectral density of random phase fluctuations is the polynomial law
\begin{align}
	S_{\varphi}(f) = \sum_{i=-n}^{0}{b_{i}f^{i}}
	\label{eqn:Sphi_powerlaw}
\end{align}
with usually $n \le 4$. $i = 0$ and $i = -1$ refer to white phase noise and $1/f$ flicker phase noise, respectively, which are the main processes in two-port components \cite[p. 23]{Rubiola.2009} \cite{Boudot.2012}. As will be shown further below, in closed-loop systems white phase noise results in white frequency noise ($i = -2$) and flicker phase noise results in flicker frequency noise ($i = -3$). Higher order effects like random walk of frequency ($i = -4$) are related to environmental changes like e.g. temperature drifts, humidity, and vibrations \cite{Rutman.1978}. All phase noise measurements were performed utilizing a \textit{Rohde \& Schwarz FSWP} phase noise analyzer.

\subsection{SAW Sensor}
\label{subsec:noise_analysis_saw_sensor}

For the measurement of the power spectral density of the random phase fluctuations of the magnetically coated SAW device $S_{\varphi}^{\mathrm{SAW}}(f)$, the sensor has been placed inside an ultra high magnetic field shielding mu-metal cylinder \textit{ZG1} from \textit{Aaronia AG} in order to avoid environmental magnetic distortions. For the sensor in its operating point (blue line in Fig.~\ref{fig:saw_phasenoise}) of ${B_{\mathrm{bias}} = 0}$ the measurement reveals a 1/$f$ flicker phase noise which can be quantified properly by ${b_{-1}/f}$ with ${b_{-1} = \SI{8e-11}{rad^2}}$. If the sensor is magnetically saturated utilizing a strong permanent magnet (green line in Fig.~\ref{fig:saw_phasenoise}) the flicker phase noise decreases by more than \SI{20}{dB}, indicating that the sensor's resolution is limited by magnetic magnetic noise, e.g. magnetic domain wall motions. The underlying mechanisms and possibilities to reduce the flicker phase noise are currently still under investigation. Although both measurements were conducted for an input power of $P_{\mathrm{SAW}} = \SI{0}{dBm}$ the white phase noise level differs because under magnetic saturation the insertion loss decreases. Note that both white phase noise floors are slightly impaired by an utilized preamplifier (\textit{ZFL-1000LN+} from \textit{Mini-Circuits} with a noise figure of ${F \approx \SI{3}{dB}}$) at the RF input of the phase noise analyzer.

\begin{figure*}[t]
	\centering	
	\begin{subfigure}[t]{0.336\textwidth}
		\centering
		\includegraphics[width=1.0\linewidth]{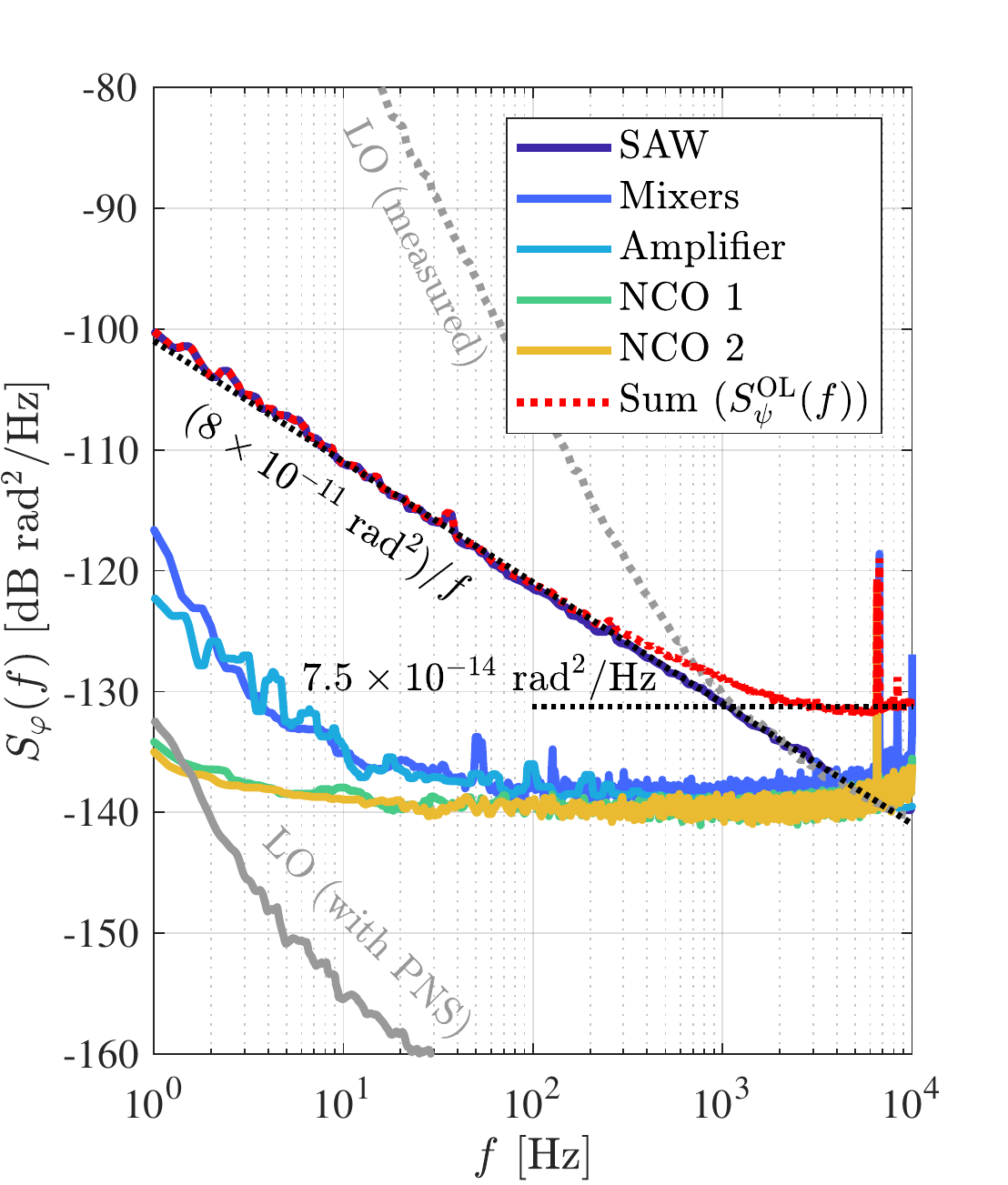}
		\caption{Measured phase noise of the individual components of the open-loop SAW sensor system together with their sum}
		\label{fig:pm_phasenoise}
	\end{subfigure}
	~
	\begin{subfigure}[t]{0.257\textwidth}
		\centering
		\includegraphics[width=1.0\linewidth]{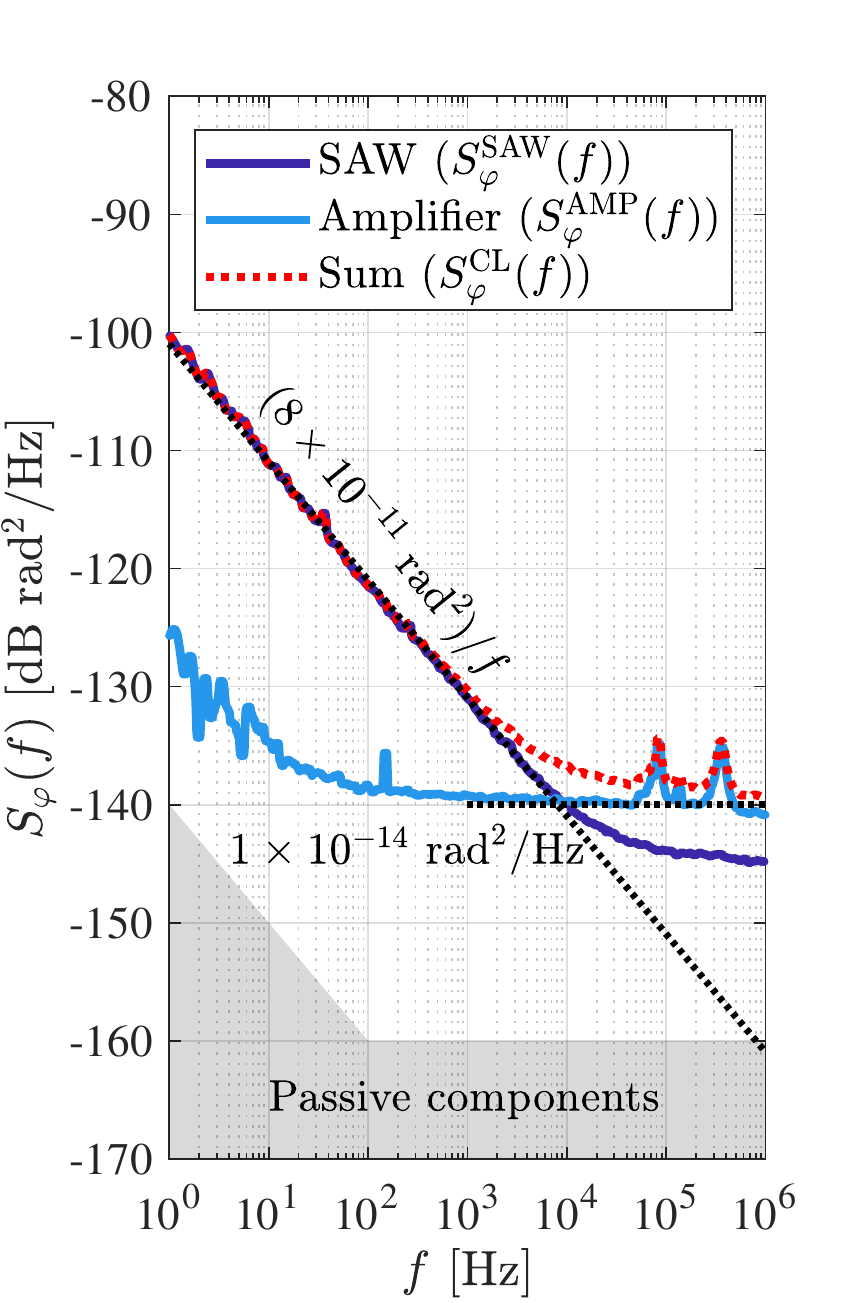}
		\caption{Measured phase noise of the individual components of the closed-loop SAW sensor system}
		\label{fig:fm_phasenoise_components}
	\end{subfigure}
	~
	\begin{subfigure}[t]{0.356\textwidth}
		\centering
		\includegraphics[width=1.0\linewidth]{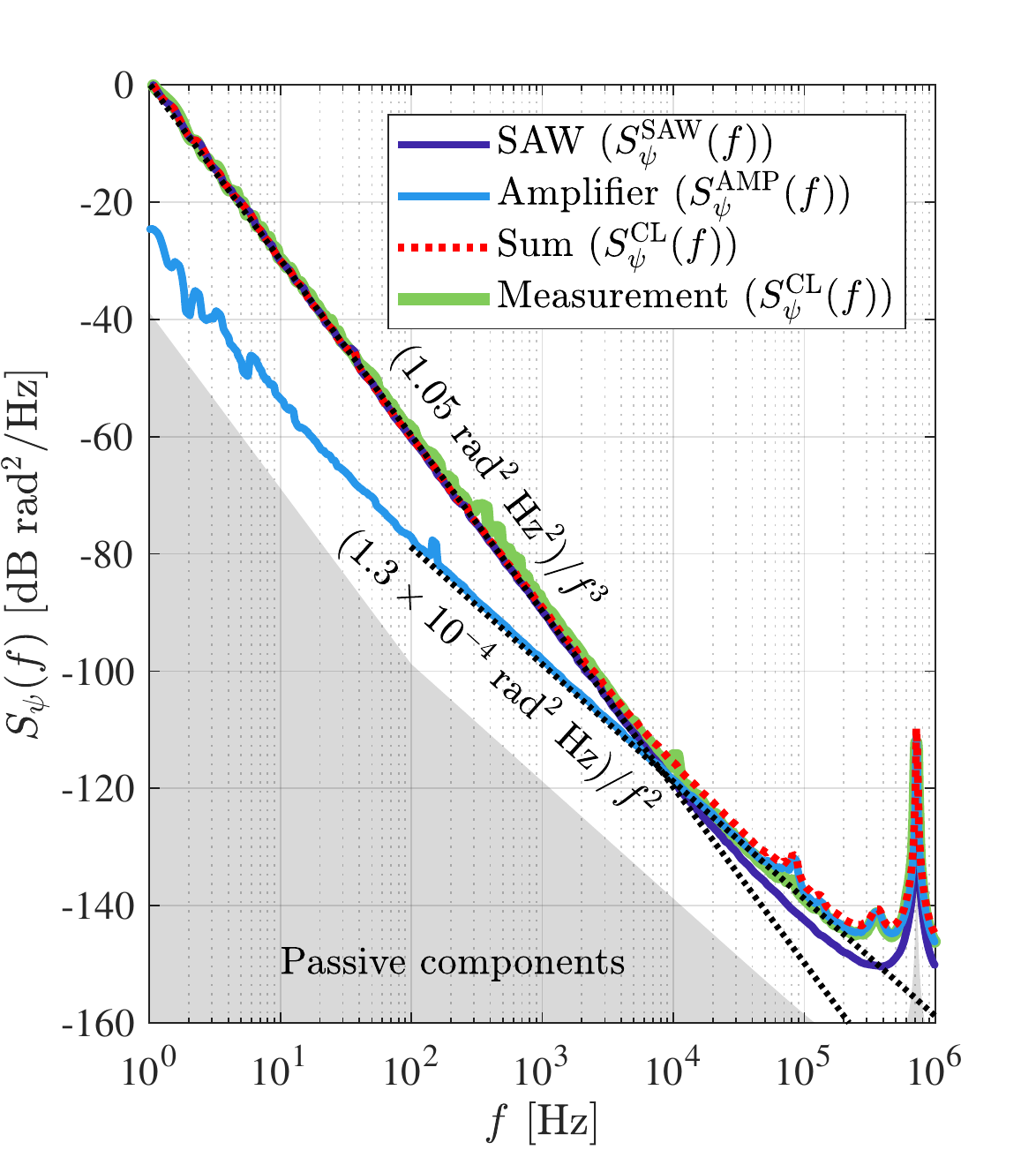}
		\caption{Calculated contribution to the oscillator's phase noise of the individual components of the closed-loop SAW sensor system together with their sum in comparison to a measurement of the oscillator's phase noise.}
		\label{fig:fm_phasenoise_oscillator}
	\end{subfigure}
	~
	\caption{Power spectral densities of the random phase fluctuations of the individual components of the open-loop sensor system (\subref{fig:pm_phasenoise}) and of the closed-loop sensor system (\subref{fig:fm_phasenoise_components}). The calculated contribution of the individual components of the closed-loop SAW sensor system to the oscillator's phase noise is in perfect agreement with a measurement of the oscillator's phase noise (\subref{fig:fm_phasenoise_oscillator}). All measurements of the individual components are performed in their operating points regarding input power, gain, and frequency exactly like depicted in Fig.~\ref{fig:readout_systems}.}
	\label{fig:pm_fm_phasenoise}
\end{figure*}

\subsection{Open-Loop System}
\label{subsec:noise_analysis_open_loop_system}

For the open-loop system, Fig.~\ref{fig:pm_phasenoise} depicts the measured phase noise contributions of the various electrical components. The high phase noise of the LO with values of \SI{-31}{dB~rad^2/Hz} at \SI{1}{Hz} and \SI{-74}{dB~rad^2/Hz} at \SI{10}{Hz} (not shown for clarity) is largely suppressed in the downconverter stage. The degree of phase noise suppression as a function of the frequency and the sensor's group delay
\begin{align}
		\mathrm{PNS}(f,\tau_{\mathrm{g}}) = 20 \log_{10}\left( 2 \sin\left(\pi f \tau_{\mathrm{g}}\right) \right) \mathrm{dB}
		\label{eqn:PNS}
\end{align}
has already been determined and was experimentally verified in previous investigations \cite{Durdaut.2018,Durdaut.2019}. Thus, the SAW sensor dominates the overall phase noise for frequencies \SI{<1}{kHz}. At $f = \SI{1}{Hz}$ the mixers and the amplifier contribute a phase noise density of approx. \SI{-120}{dB~rad^2/Hz}. With values \SI{<-135}{dB~rad^2/Hz}, the phase noise of the NCOs, i.e. the in-phase and quadrature (IQ) signals are negligible at low frequencies. However, for frequencies \SI{>1}{kHz} the sum of the phase noise contributions of the various electrical components become dominant and result in a white phase noise level of \SI{7.5e-14}{rad^2/Hz}.

The LOD of the open-loop magnetic field sensor system in units of $\mathrm{T}/\sqrt{\mathrm{Hz}}$ can be calculated \cite{Durdaut.2019} by 
\begin{align}
		\mathrm{LOD}_{\mathrm{PM}}(f) = \frac{\sqrt{S_{\psi}^{\mathrm{OL}}(f)}}{\mathcal{S}_{\mathrm{PM}}},
		\label{eqn:LODPM}
\end{align}
where $S_{\psi}^{\mathrm{OL}}(f)$ is the power spectral density of the fluctuating phase $\psi_{\mathrm{OL}}(t)$ introduced in Eq.~\eqref{eqn:sPMt}. This power spectral density is the sum of both, the phase noise of the SAW sensor and the phase noise contributions of all other electrical components of the open-loop system (mixers, amplifier, NCOs, and LO with PNS) shown in Fig.~\ref{fig:pm_phasenoise}. Please note that Eq.~\eqref{eqn:PNS} and Eq.~\eqref{eqn:LODPM} are simplified expressions for delay line sensors offering bandwidths higher than the frequency to be analyzed, which is true for the sensors under investigation. More accurate expressions for narrow band delay line sensors can be found in \cite{Durdaut.2019}.

\subsection{Closed-Loop System}
\label{subsec:noise_analysis_closed_loop_system}

For the noise analysis of the closed-loop system only the contributions of the electrical components inside the oscillator loop are relevant. In fact, these are only the phase noise of the SAW sensor and the amplifier. The phase noise spectra of the limiter and of the directional coupler are actually not measurable with the \textit{Rohde \& Schwarz FSWP} phase noise analyzer in a reasonable time because values as low as approx. \SI{-180}{dB~rad^2/Hz} at \SI{10}{Hz} can be typically found for such passive devices \cite{Rubiola.2002}. In addition to the intrinsic noise of the SAW sensor, Fig.~\ref{fig:fm_phasenoise_components} shows the measured phase noise of the amplifier and a rough estimation of the contributions of the passive components utilized inside the oscillator loop. Both the noise contributions of the passive components and the amplifier are negligible for $f < \SI{8}{kHz}$. Actually, the influence of the amplifier above this frequency could easily be further decreased by utilizing a fixed-gain amplifier instead of a variable-gain amplifier used here. In order to calculate the LOD for the closed-loop sensor system it is necessary to convert the phase noise of the individual components into oscillator phase noise at the output of the oscillator, i.e. at the output of the directional coupler. Following the derivations in \cite{Durdaut.2019}, random phase fluctuations $\varphi(t)$ fed into an ideal delay line oscillator result in oscillator phase noise $\psi(t)$ according to the transfer function
\begin{align}
		|H(f)|^2 &= \left|\frac{\Psi(f)}{\Phi(f)}\right|^2 = \frac{1}{\left|1 - e^{-j 2 \pi f \tau_{\mathrm{g}}}\right|^2}\notag\\
		&= \frac{1}{2~(1-\cos(2 \pi f \tau_{\mathrm{g}}))} = \frac{1}{4 \sin^2(\pi f \tau_{\mathrm{g}})},
		\label{eqn:PsidurchPhisq}
\end{align}
where $\Phi(f)$ and $\Psi(f)$ represent the Fourier transforms of the fluctuating phases $\varphi(t)$ and $\psi(t)$. A more general representation of this transfer function which also takes into account the frequency-selective characteristic of narrow band delay line sensors is given in \cite{Durdaut.2019}. However, the SAW sensor under investigation offers a relatively high \SI{-3}{dB} bandwidth of $\Delta f = \SI{4.6}{MHz}$, such that the approximation in Eq.~\eqref{eqn:PsidurchPhisq} is valid for the frequency range of up to \SI{1}{MHz} analyzed in the following (Fig.~\ref{fig:pm_fm_phasenoise} and Fig.~\ref{fig:lod}). The power spectral density of the SAW delay line oscillator's overall random phase fluctuations is then given by
\begin{align}
		S_{\psi}^{\mathrm{CL}}(f) = |H(f)|^2 ~ S_{\varphi}^{\mathrm{CL}}(f).
		\label{eqn:SpsifvonHfsqSphif}
\end{align}
The sum of the phase noise contributions of the SAW sensor, the amplifier, and the (negligible) passive components ${S_{\varphi}^{\mathrm{CL}}(f) = S_{\varphi}^{\mathrm{SAW}}(f) + S_{\varphi}^{\mathrm{AMP}}(f)}$ shown in Fig.~\ref{fig:fm_phasenoise_components} are converted into oscillator phase noise ${S_{\psi}^{\mathrm{CL}}(f) = S_{\psi}^{\mathrm{SAW}}(f) + S_{\psi}^{\mathrm{AMP}}(f)}$ using Eq.~\eqref{eqn:SpsifvonHfsqSphif} and are shown in Fig.~\ref{fig:fm_phasenoise_oscillator}. As one would expect, the SAW sensor is still contributing the dominant noise for $f < \SI{8}{kHz}$ whereas the overall noise floor is limited by the amplifier's noise for $f > \SI{8}{kHz}$. The sum of the individual contributions (dotted line) perfectly agrees with a direct measurement of the oscillator's phase noise, and thus confirms the transfer function $|H(f)|^2$ from Eq.~\eqref{eqn:PsidurchPhisq}. Even the predicted increase in oscillator phase noise at ${f = 1/\tau_{\mathrm{g}} \approx \SI{720}{kHz}}$, due to the pole of the transfer function, perfectly matches the measurement. With the relation between the power spectral density of arbitrary random phase fluctuations $S_{\varphi}(f)$ and the power spectral density of frequency fluctuations $S_{\Delta f_0}(f) = f^2 S_{\varphi}(f)$ \cite{Rutman.1978} in units of $\mathrm{Hz}^2/\mathrm{Hz}$ and for a carrier signal of frequency $f_0$, the LOD of the closed-loop magnetic field sensor system in units of $\mathrm{T}/\sqrt{\mathrm{Hz}}$ can be calculated \cite{Durdaut.2019} by 
\begin{align}
		\mathrm{LOD}_{\mathrm{FM}}(f) &= \frac{\sqrt{S_{\Delta f_0}^{\mathrm{CL}}(f)}}{\mathcal{S}_{\mathrm{FM}}} = \frac{\sqrt{f^2 ~ S_{\psi}^{\mathrm{CL}}(f)}}{\mathcal{S}_{\mathrm{FM}}}\notag\\
		&= \frac{f ~ |H(f)| ~ \sqrt{S_{\varphi}^{\mathrm{CL}}(f)}}{\mathcal{S}_{\mathrm{FM}}}.
		\label{eqn:LODFM}
\end{align}

\subsection{Comparison of the Limit of Detection (LOD)}
\label{subsec:noise_analysis_comparison_lod}

\begin{figure}[t]
	\centering
	\includegraphics[width=0.5\textwidth]{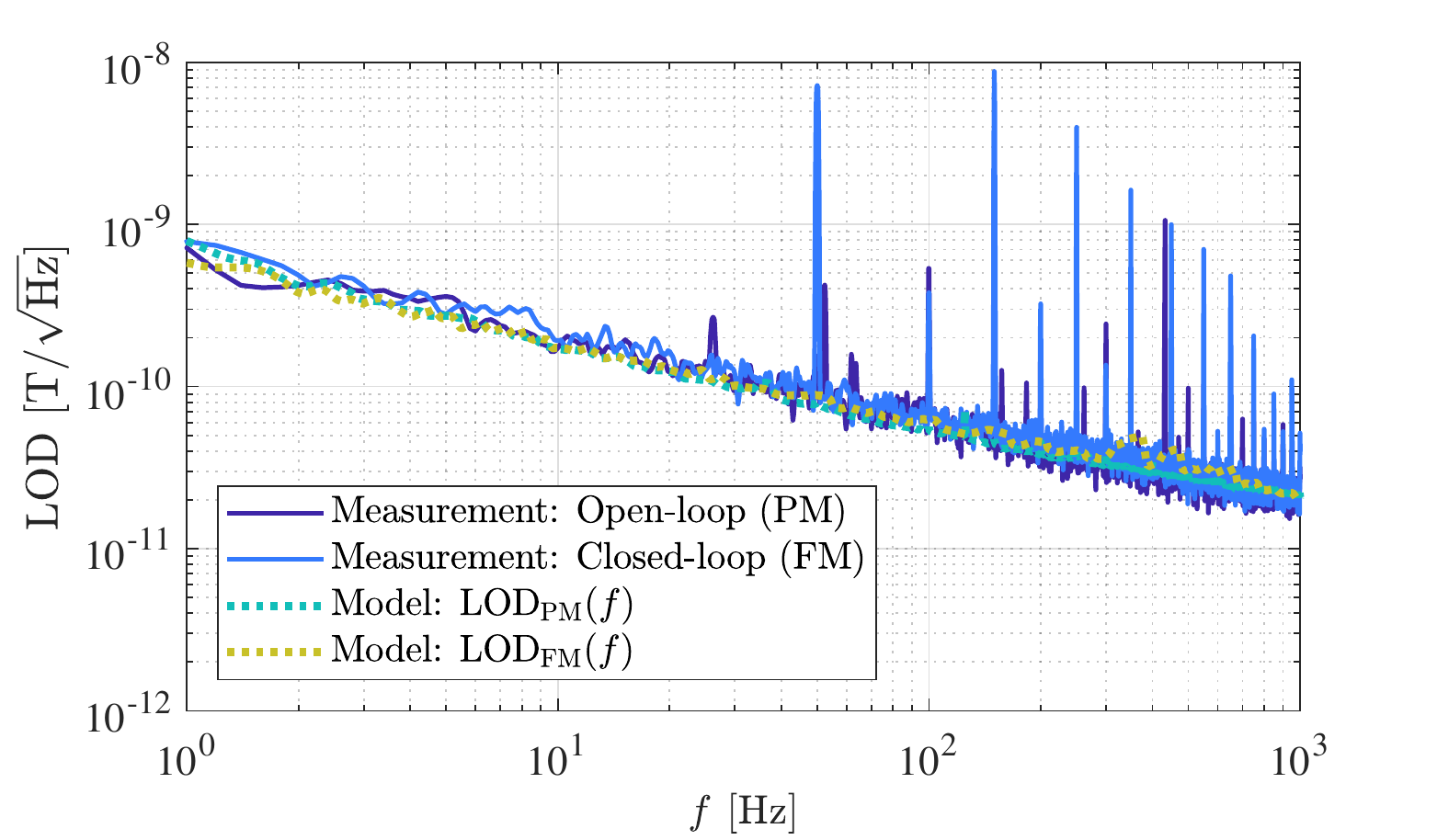}
	\caption{Measured limits of detection (LOD) utilizing the open-loop and closed-loop readout systems presented in Sec.~\ref{sec:readout_systems} compared to their predictions according to Eq.~\eqref{eqn:LODPM} and \eqref{eqn:LODFM}. Both operational modes result in the same LOD. Spurious signals are due to the power supply of the amplifiers and can be disregarded.}
	\label{fig:lod}
\end{figure}

Both expressions for the LOD in the open-loop and for the closed-loop sensor system (Eq.~\eqref{eqn:LODPM} and \eqref{eqn:LODFM}) directly scale with the amplitude spectral densities of the randomly fluctuating phase of the SAW sensor and the readout electronics. Thus, depending on how carefully the readout systems are designed in terms of phase noise, the LOD can differ. However, as shown for both systems the SAW sensor contributes the dominant phase noise at least for frequencies \SI{<1}{kHz}. Thus, when assuming that the SAW sensor contributes the dominant phase noise in each system, equality between $\mathrm{LOD}_{\mathrm{FM}}(f)$ and $\mathrm{LOD}_{\mathrm{PM}}(f)$ can easily be shown. With Eq.~\eqref{eqn:PsidurchPhisq} and for frequencies $f \ll \Delta f$ Eq.~\eqref{eqn:LODFM} can be written as
\begin{align}
		\mathrm{LOD}_{\mathrm{FM}}(f) \approx \frac{f ~ \sqrt{S_{\varphi}^{\mathrm{SAW}}(f)}}{2 \sin(\pi f \tau_{\mathrm{g}}) ~ \mathcal{S}_{\mathrm{FM}}}.
		\label{eqn:LODFM1}
\end{align}
Replacing both the sine with a small-angle approximation (${\sin(x) \approx x}$) which is valid for ${f \tau \ll 0.1}$ and replacing $\mathcal{S}_{\mathrm{FM}}$ with the previously defined closed-loop sensitivity ${\mathcal{S}_{\mathrm{FM}} = \mathcal{S}_{\mathrm{mag}}/\mathcal{S}_{\mathrm{elec}} = \mathcal{S}_{\mathrm{mag}}/(2 \pi \tau_{\mathrm{g}})}$, Eq.~\eqref{eqn:LODFM1} yields
\begin{align}
		\mathrm{LOD}_{\mathrm{FM}}(f) \approx \frac{2 \pi f \tau_{\mathrm{g}} ~ \sqrt{S_{\varphi}^{\mathrm{SAW}}(f)}}{2 \pi f \tau_{\mathrm{g}} ~ \mathcal{S}_{\mathrm{mag}}}.
		\label{eqn:LODFM2}
\end{align}
With the definition of the open-loop sensitivity ${\mathcal{S}_{\mathrm{PM}} = \mathcal{S}_{\mathrm{mag}}}$, Eq.~\eqref{eqn:LODFM2} then virtually equals the LOD of the open-loop system
\begin{align}
		\mathrm{LOD}_{\mathrm{FM}}(f) \approx \frac{\sqrt{S_{\varphi}^{\mathrm{SAW}}(f)}}{\mathcal{S}_{\mathrm{mag}}} = \mathrm{LOD}_{\mathrm{PM}}(f).
		\label{eqn:LODeq}
\end{align}
For both sensor systems presented in Sec.~\ref{sec:readout_systems} the LOD was measured as described in \cite{Kittmann.2018} and compared with calculations based on Eq.~\eqref{eqn:LODPM} and \eqref{eqn:LODFM}. The results are shown in Fig.~\ref{fig:lod} and reveal that the measured LODs not only agree with their according predictions but also that the results are the same for open-loop (PM) and closed-loop (FM) operation for frequencies \SI{<1}{kHz}. A value of approx. $\SI{170}{\mathrm{pT}/\sqrt{\mathrm{Hz}}}$ is reached at a frequency of \SI{10}{Hz}.

\subsection{Time Domain Uncertainty}
\label{subsec:noise_analysis_time_domain_uncertainty}

The output of a sensor system is most often exploited as a stream of data. Designing a system, it is wise to set the averaging time $\tau$ (please do not mistake $\tau$ with the phase delay $\tau_{\varphi}$ or the group delay $\tau_{\mathrm{g}}$) equal to the sampling interval. This is an efficient use of the information because there is no overlap between measurements, and no dead time. A longer $\tau$ results in higher rejection of white noise, and in turn a lower uncertainty, at the cost of a slower data rate. The question about the optimum measurement time $\tau_{\mathrm{opt}}$, beyond which the uncertainty no longer improves, is best answered describing the fluctuations in terms of the \textit{Allan variance} (AVAR) \cite{Barnes.1971,Stein.2010}. The Allan variance of an arbitrary quantity $y(t)$, denoted by $\sigma_{y}^2(\tau)$, is an extension of the regular variance that makes the measurement time $\tau$ appear explicitly, and also converges in the presence of flicker noise, random walk and drift. Describing random fluctuations of $y(t)$ by the power spectral density ${S_y(f) = h_{-1}/f + h_0}$, it holds that ${\sigma_{y}^2(\tau) = 2 \ln(2) h_{-1} + h_0/(2 \tau)}$. We have shown in \cite{Durdaut.2019} that the optimum measurement time is
\begin{align}
		\tau_{\mathrm{opt}} = \frac{1}{4 \ln(2)} \frac{1}{f_{\mathrm{c}}} \approx 0.36 \frac{1}{f_{\mathrm{c}}}
\end{align}
where ${f_{\mathrm{c}} = h_{-1}/h_{0}}$ is the corner frequency where the flicker noise crosses the white noise.

For the open-loop system, we identify the quantity $y(t)$ with the random phase $\varphi(t)$, thus ${h_i = b_i}$. With a flicker phase noise of ${b_{-1} = \SI{8e-11}{rad^2}}$ and white phase noise of ${b_{0} = \SI{7.5e-14}{rad^2/Hz}}$ (Fig.~\ref{fig:pm_phasenoise}) the crossover frequency yields
\begin{align}
		f_{\mathrm{c}}^{\mathrm{OL}} = \frac{h_{-1}}{h_{0}} = \frac{b_{-1}}{b_{0}} = \SI{1.07}{kHz}.
\end{align}

For the closed-loop system, we identify the quantity $y(t)$ with the fractional frequency fluctuation $(\Delta f_0)(t)/f_0$, thus ${h_{-1} = b_{-3}/f_0^2}$ and ${h_{0} = b_{-2}/f_0^2}$. With a flicker frequency noise ${b_{-3} = \SI{1.05}{rad^2 Hz^2}}$ and white frequency noise of ${b_{-2} = \SI{1.3e-4}{rad^2 Hz}}$ (Fig.~\ref{fig:fm_phasenoise_oscillator}) the crossover frequency yields
\begin{align}
		f_{\mathrm{c}}^{\mathrm{CL}} = \frac{h_{-1}}{h_{0}} = \frac{b_{-3}}{b_{-2}} = \SI{8.08}{kHz}.
\end{align}

Thus, for the presented readout systems, the optimum measurement times differ and result in ${\tau_{\mathrm{opt}}^{\mathrm{OL}} \approx \SI{338}{\mu s}}$ and ${\tau_{\mathrm{opt}}^{\mathrm{CL}} \approx \SI{45}{\mu s}}$. The equivalent magnetic uncertainties then yield ${\sigma_y^{\mathrm{OL}}(\tau_{\mathrm{opt}}^{\mathrm{OL}})/\mathcal{S}_{\mathrm{PM}} \approx \SI{903}{pT}}$ for the open-loop system and ${\sigma_y^{\mathrm{CL}}(\tau_{\mathrm{opt}}^{\mathrm{CL}}) f_0 / \mathcal{S}_{\mathrm{FM}} \approx \SI{904}{pT}}$ for the closed-loop system. For ideal sensor systems, in which the sensor contributes the dominant phase noise for all frequencies, $\tau_{\mathrm{opt}}$ would be the same for both readout systems \cite{Durdaut.2019}.

\section{Conclusion}
\label{sec:conclusion}
Based on the electrical and magnetic properties of a SAW delay line magnetic field sensor low noise readout systems for the detection of the magnetic signal in an open-loop and closed-loop configuration, respectively, are presented and compared. By analyzing the phase noise contributions of the individual electrical components of each readout system it is revealed that the sensor itself contributes the dominant phase noise for a wide range of frequencies. In particular, it is shown that the phase noise of the SAW delay line oscillator, i.e. the closed-loop system, can be accurately predicted such that according expressions for the calculation of the limits of detection can be derived. Based on these equations equality between the LOD of open-loop and closed-loop SAW delay line readout can be shown even analytically assuming that the sensor contributes the dominant phase noise. This equality is verified by according measurements. These results are applicable to all kinds of phase sensitive delay line sensors. Therefore, the decision-making process for selecting a certain readout structure for a given sensor should mainly be based on the possibility to reduce the electronic's phase noise contribution. However, only open-loop systems allow for the characterization of the sensor's transmission properties as well as to identify the optimum operating point.

\section*{Acknowledgment}
This work was supported (1) by the German Research Foundation (Deutsche Forschungsgemeinschaft, DFG) through the Collaborative Research Centre CRC 1261 \textit{Magnetoelectric Sensors: From Composite Materials to Biomagnetic Diagnostics}, (2) by the ANR Programme d'Investissement d'Avenir (PIA) under the Oscillator IMP project and the FIRST-TF network, and (3) by grants from the R\'{e}gion Bourgogne Franche-Comt\'{e} intended to support the PIA. In addition, the authors would like to thank Wolfgang Taute for fruitful discussions and practical help during the development of the readout electronics.

\ifCLASSOPTIONcaptionsoff
  \newpage
\fi




\bibliographystyle{IEEEtran}
\bibliography{mybibfile}


%

\begin{IEEEbiography}[{\includegraphics[width=1in,height=1.25in,clip,keepaspectratio]{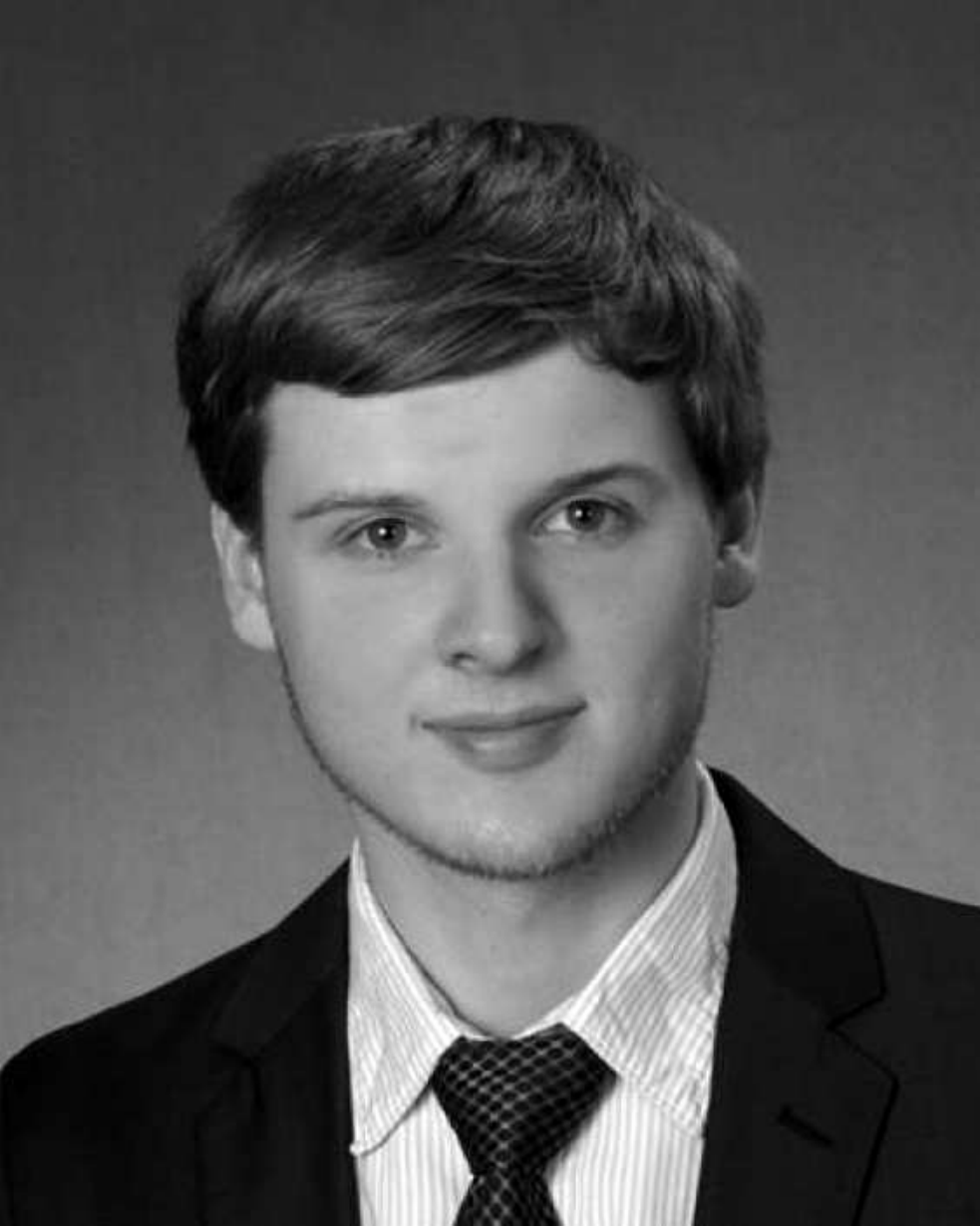}}]{Phillip Durdaut}
(S'17) received a B.Eng. degree in Information and Electrical Engineering from the University of Applied Sciences Hamburg, Hamburg, Germany, in 2013 and an M.Sc. degree in Electrical Engineering and Information Technology from Kiel University, Kiel, Germany, in 2015. He is currently working towards a Dr.-Ing. degree at the Chair of Microwave Engineering, Kiel University. His main research interests include electronics with maximum signal-to-noise ratio for sensor systems based on thin-film magnetoelectric- and SAW sensors.
\end{IEEEbiography}

\begin{IEEEbiography}[{\includegraphics[width=1in,height=1.25in,clip,keepaspectratio]{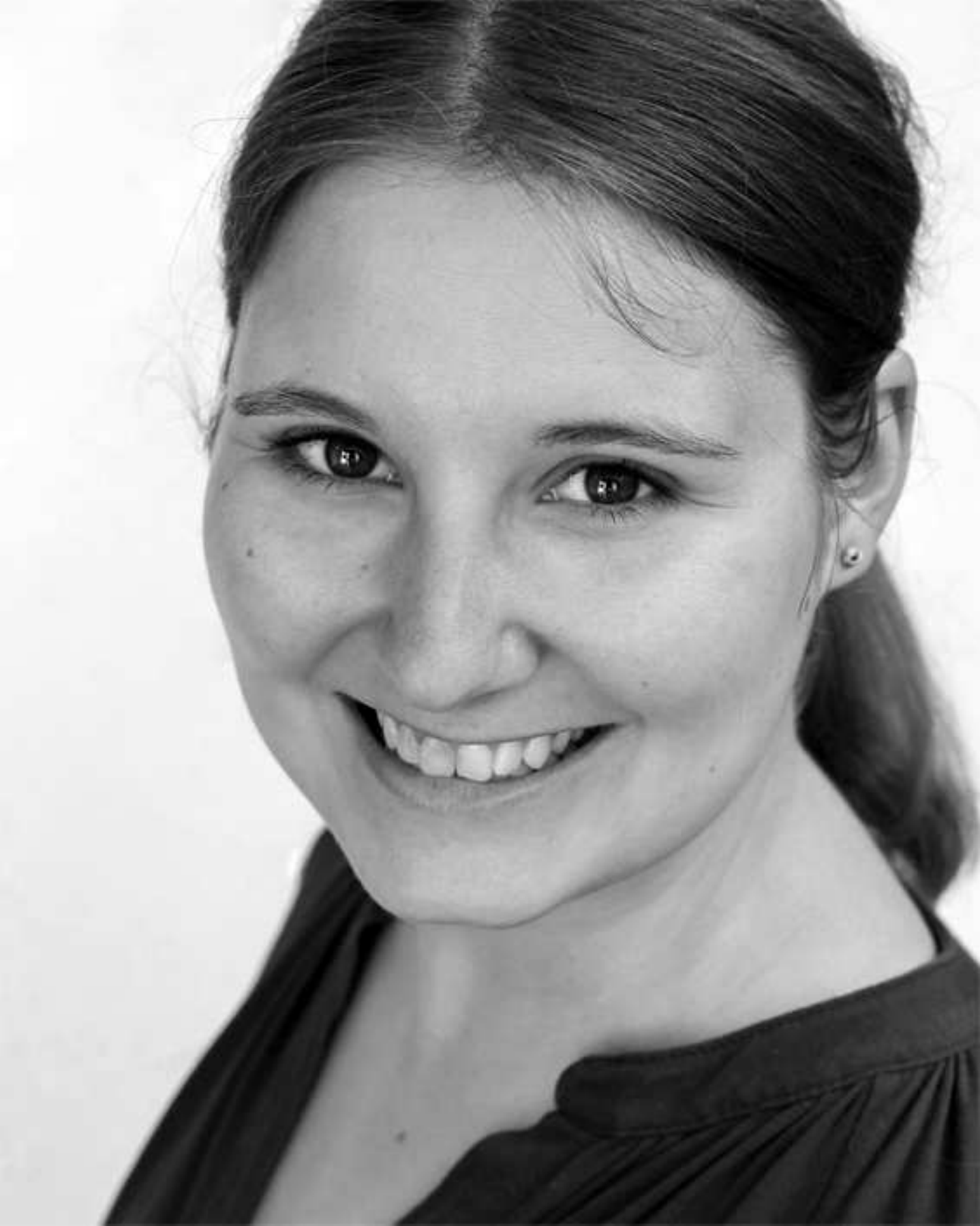}}]{Anne Kittmann}
received her B.Sc. degree and M.Sc. degree in Material Science and Engineering from the Kiel University, Kiel, Germany, in 2012 and 2015, respectively. During her current work towards a Dr.-Ing. degree at the Chair of Inorganic Functional Materials, Kiel University, she is working on the development of sensors for magnetic field measurement. The research focus is on thin-film magnetoelectric- and surface acoustic wave sensors with high sensitivity and lowest limit of detection.  
\end{IEEEbiography}

\begin{IEEEbiography}[{\includegraphics[width=1in,height=1.25in,clip,keepaspectratio]{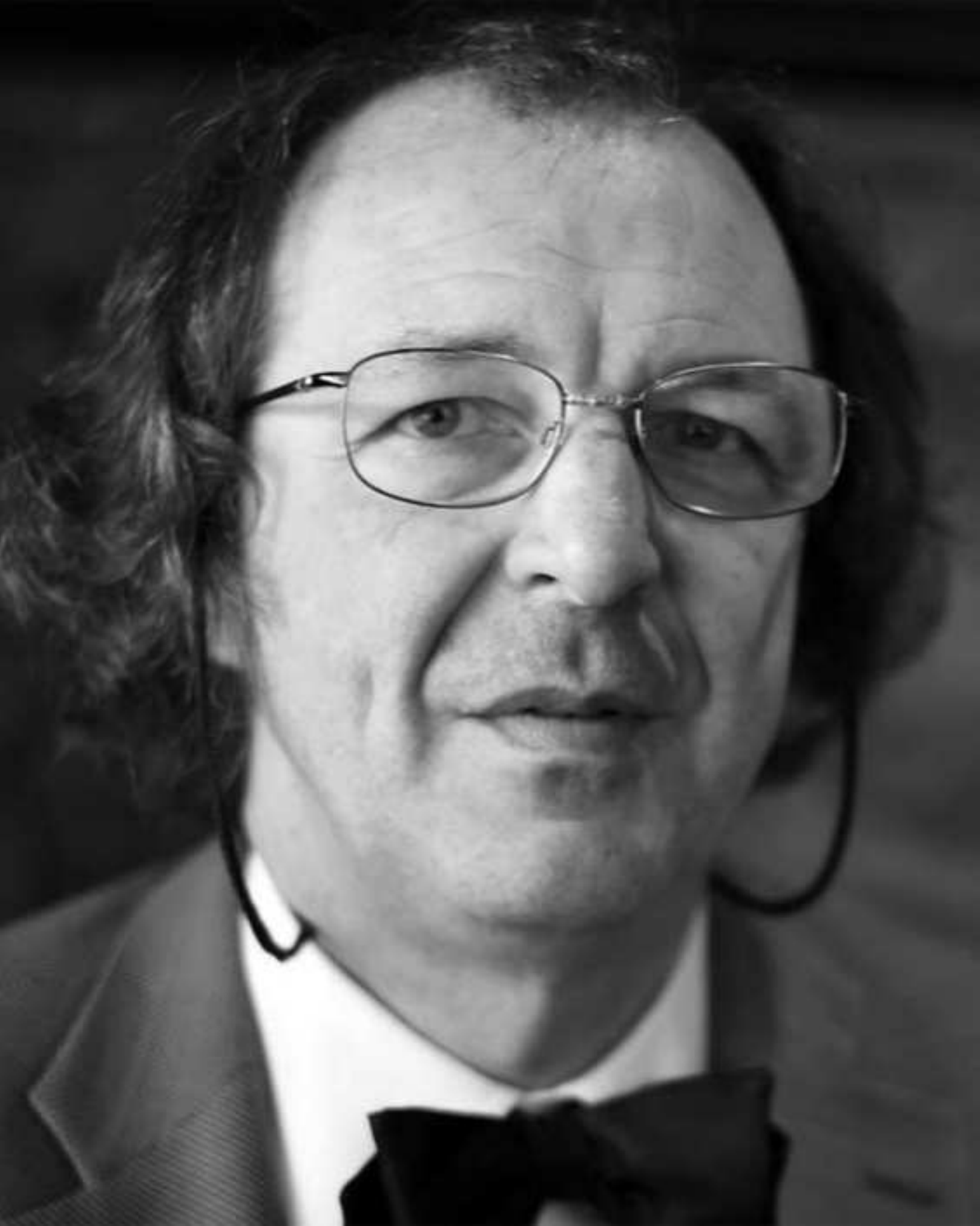}}]{Enrico Rubiola}
(M'04) received the M.S. degree in electronics engineering from the Politecnico di Torino, Turin, Italy, in 1983, the Ph.D. degree in metrology from the Italian Ministry of Scientific Research, Rome, Italy, in 1989, and the D.Sc. degree in time and frequency metrology from the University of Franche-Comt\'{e} (UFC), Besan\c{c}on, France, in 1999. He has been a Researcher with the Politecnico di Torino, a Professor with the University of Parma, Parma, Italy, and Universit\'{e} Henri Poincar\'{e}, Nancy, France, and a Guest Scientist with the NASA Jet Propulsion Laboratory, Pasadena, CA, USA. He has been a Professor with UFC and a Scientist with the Centre National de la Recherche Scientifique, Franche Comt\'{e} \'{E}lectronique M\'{e}canique Thermique et Optique-Sciences et Technologies Institute, Besan\c{c}on, since 2005, where he currently serves as the Deputy Director of the Time and Frequency Department. He has also investigated various topics of electronics and metrology, namely, navigation systems, time and frequency comparisons, atomic frequency standards, and gravity. He is the PI of the Oscillator IMP project, a platform for the measurement of short-term frequency stability and spectral purity. His current research interests include precision electronics, phase-noise, amplitude-noise, frequency stability and synthesis, and low-noise oscillators from the low RF region to optics.
\end{IEEEbiography}

\begin{IEEEbiography}[{\includegraphics[width=1in,height=1.25in,clip,keepaspectratio]{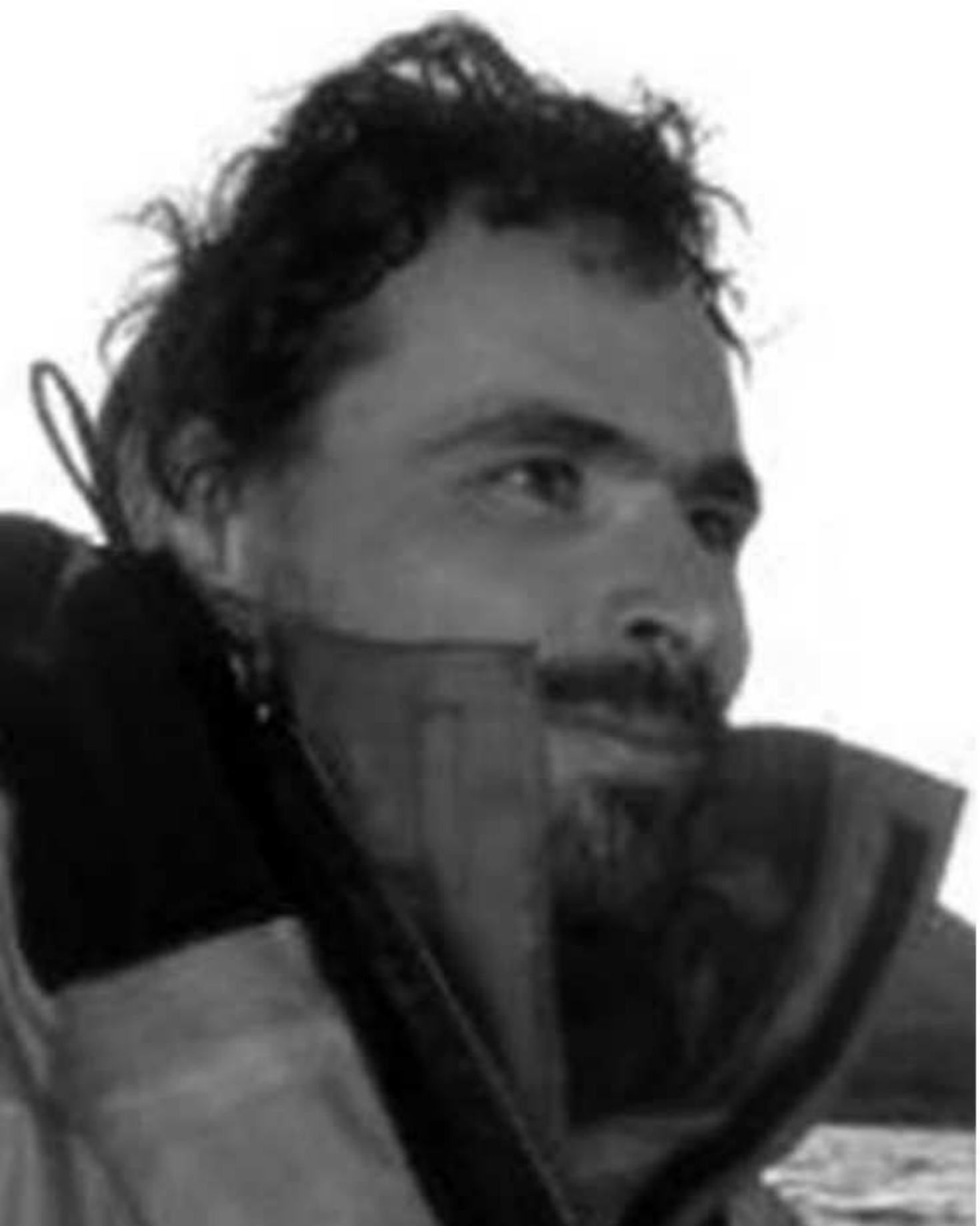}}]{Jean-Michel Friedt}
obtained his Ph.D. degree in 2000. He worked as a postdoctoral fellow in IMEC, Leuven, Belgium, on acoustic and optical biosensors for characterizing organic thin film properties before joining the group of S. Ballandras in 2004 at FEMTO-ST, Besan\c{c}on, France, for the development of passive sensors interrogated through a wireless link. He has been an engineer with the company SENSeOR since its creation in 2006, hosted by the Time and Frequency Department of FEMTO-ST. His interests include scanning probe microscopy, passive radio-frequency sensors and the associated radar-like electronics implemented as software-defined radio, and the combination of optical and acoustic methods for characterizing thin films.
\end{IEEEbiography}

\begin{IEEEbiography}[{\includegraphics[width=1in,height=1.25in,clip,keepaspectratio]{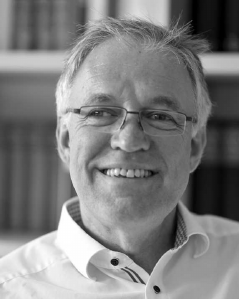}}]{Eckhard Quandt}
received his Diploma and Dr.-Ing. degrees in physics from the Technische Universit\"at Berlin, Germany, in 1986 and 1990, respectively. He is currently a Professor with Kiel University, Kiel, Germany, where he is the Director at the Institute for Materials Science. He is the spokesperson for the DFG CRC 1261 Magnetoelectric Sensors: From Composite Materials to Biomagnetic Diagnostics, and is a member of Acatech, National Academy of Science and Engineering. His scientific focus is material research on smart materials and multiferroics and the development of sensors and actuators for microelectromechanical systems and nanoelectromechanical systems using such materials.
\end{IEEEbiography}

\begin{IEEEbiography}[{\includegraphics[width=1in,height=1.25in,clip,keepaspectratio]{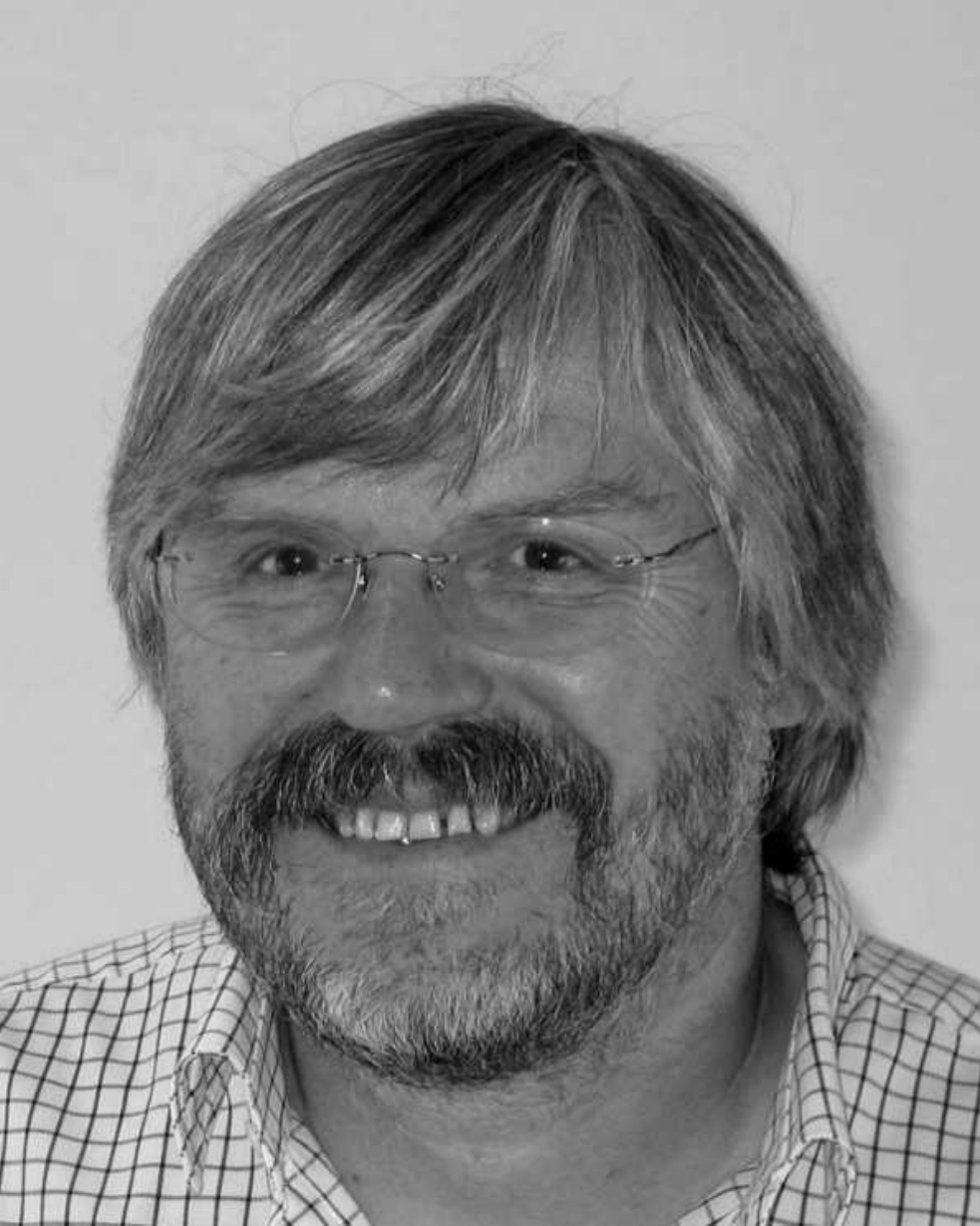}}]{Reinhard Kn\"ochel}
(M'88-SM'90-F'08-LF'18) received a Dipl.-Ing. degree in electrical engineering and a Dr.-Ing. degree from the Technical University, Braunschweig, Germany, in 1975 and 1980, respectively. From 1980 to 1986, he was a Principal Scientist with the Philips Research Laboratory, Hamburg, Germany. In 1986, he joined the Hamburg University of Technology, where he was a Full Professor of Microwave Electronics until 1993. Between 1993 and 2013, he held the Chair in Microwave Engineering Kiel University, Kiel, Germany. His research interests include magnetic field sensors, active and passive microwave components, ultrawideband technology, microwave measurement techniques, industrial microwave sensors and radar.
\end{IEEEbiography}

\begin{IEEEbiography}[{\includegraphics[width=1in,height=1.25in,clip,keepaspectratio]{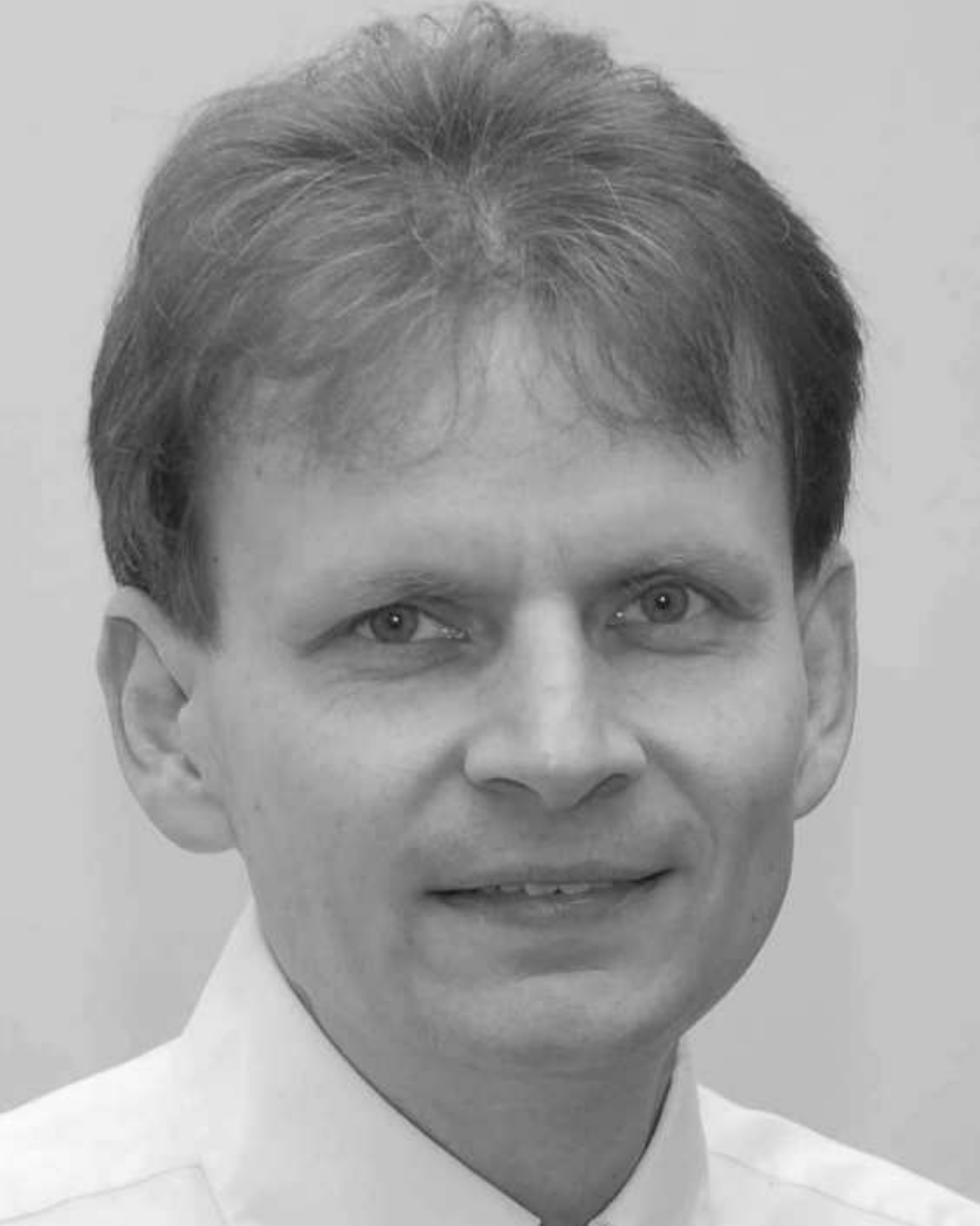}}]{Michael H\"oft}
(M'04-SM'08) received a Dipl.-Ing. degree in electrical engineering and a Dr.-Ing. degree from the Hamburg University of Technology, Hamburg, Germany, in 1997 and 2002, respectively. From 2002 to 2013, he was with the Communications Laboratory, European Technology Center, Panasonic Industrial Devices Europe GmbH, L\"uneburg, Germany. He was a Research Engineer and then a Team Leader, where he was involved in the research and development of microwave circuitry and components, particularly filters for cellular radio communications. From 2010 to 2013, he was also a Group Leader for the research and development of sensor and network devices. Since 2013, he has been a Full Professor at Kiel University, Kiel, Germany, in the Faculty of Engineering, where he is the Head of the Chair of Microwave Engineering of the Institute of Electrical Engineering and Information Technology. His research interests include active and passive microwave components, (sub-) millimeter-wave quasi-optical techniques and circuitry, microwave and field measurement techniques, microwave filters, microwave sensors and magnetic field sensors. He is a member of the European Microwave Association, the Association of German Engineers and a member of the German Institute of Electrical Engineers.
\end{IEEEbiography}






\vfill


\end{document}